\documentclass[11pt,a4paper]{article}

\usepackage[margin=1in]{geometry}
\usepackage{amsmath,amssymb,amsthm}
\usepackage{booktabs}
\usepackage{graphicx}
\usepackage{hyperref}
\hypersetup{hidelinks}
\usepackage{natbib}
\usepackage{tikz}
\usetikzlibrary{arrows.meta,positioning,fit,calc}
\usepackage{xcolor}
\usepackage{microtype}
\usepackage{enumitem}
\usepackage{placeins}
\usepackage{flafter}
\usepackage{float}
\usepackage{listings}
\title{Universal BCI Personalization:\\
One API for Frozen EEG Trunks and Foundation Models}
\author{
  Sergey Musienko\\
  \texttt{sergey@nimbusbci.com}
}
\date{}

\begin{document}
\maketitle

\begin{abstract}
Frozen EEG encoders proliferate; per-model fine-tune defaults do not scale.
We present \textbf{Nimbus Personalizer}: one contract
\texttt{encode}$\to$ Bayesian head \(\to\) \texttt{BrainState}
(optional affine mid-tier) that sits on heterogeneous frozen trunks without
a new personalization stack per architecture.
\textbf{Thesis (systems):} the contribution is the trunk-agnostic API---not
LDA-on-embeddings as an ML novelty---so OEMs integrate once and swap trunks.
\textbf{Evidence:} the same surface runs on five classical trunks
\{EEGNet, Shallow, Deep, Conformer, ATCNet\} \(\times\) four MI datasets
(18 cells) and on a foundation encoder (REVE) under the same Personalizer.
Where embedding capacity exists, the head is a cheap default mid-point
versus warm-start fine-tune or PEFT, costing orders of magnitude less
adaptation wall time while recovering much of the fine-tune accuracy gain;
calibration-only-when-clean holds in \(12/18\) cells.
Head gains are \emph{supporting} evidence that the API is useful where
capacity exists. Subject-level confidence intervals identify the clearest
dataset and span zero elsewhere. All results are exploratory (subject-level
bootstrap, no confirmatory tests);
the decision logic for when to escalate adaptation is addressed in our
companion work on the control layer.
\end{abstract}

\section{Introduction}
\label{sec:intro}

Deep MI encoders are no longer a single architecture:
EEGNet~\citep{lawhern2018eegnet}, Shallow/Deep
ConvNets~\citep{schirrmeister2017deep}, EEG
Conformer~\citep{song2022eegconformer}, and ATCNet~\citep{altaheri2023atcnet}
(among others) all ship as plausible frozen trunks.
Session and subject shift remain central failure
modes~\citep{lotte2018review,jayaram2016transfer}.
Our controlled stress here is a synthetic channel-mix stress grid that
induces channel mixing and bias (Section~\ref{sec:protocol})---a
proxy for recoverable distribution change, not a claim that we have
already covered every clinical shift type.
The industrial reflex---fine-tune each trunk per user---does not scale as
the set of available trunks grows.

We ask a \emph{systems} question:

\begin{quote}
\emph{Can one personalization API sit on many frozen EEG trunks without
per-model fine-tune defaults---and where is that shared surface worth
using?}
\end{quote}

We answer with \textbf{Nimbus Personalizer}: wrap a frozen
\texttt{encode} function, fit a Bayesian head (default LDA) on
calibration embeddings, optionally apply a mid-tier affine recovery map,
and expose predictions as \texttt{BrainState}.
The contribution is not ``LDA on frozen features''---that recipe is
classical---but a \emph{trunk-agnostic personalization contract} that
makes cheap head adaptation the default integration path across
architectures, with encoder fine-tune / PEFT kept as optional, priced
escape hatches.

\paragraph{Relation to companion work.}
This paper and its companion~\citep{musienko2026ladder} form a diptych.
The companion establishes the control logic and gating mechanisms for
\emph{when} to spend labels or escalate adaptation levels; the present
work focuses on the architectural framework required to \emph{execute}
that logic across diverse frozen models---the trunk-agnostic abstraction
that makes a single personalization path deployable. The contribution
here is that shared surface itself, and the accuracy--cost cells show
where it earns its keep.

\paragraph{Empirical questions.}
\begin{enumerate}[leftmargin=1.4em,itemsep=0.2em]
\item \textbf{(Primary)} Does one Personalizer API run on heterogeneous
frozen trunks (classical + foundation) without per-trunk redesign?
\item Where does the Bayesian head improve utility under shift
(supporting: capacity-dependent usefulness)?
\item Is that cheap path competitive with fine-tune / PEFT on cost
(EEGNet classical; REVE foundation, including LoRA)?
\item How often does soft cal-only-when-clean transfer, versus a strict
ordinal mid-tier schedule?
\end{enumerate}

\paragraph{Main takeaway.}
\textbf{Ship one Personalizer API; treat fine-tune / PEFT as priced
escape hatches.}
The interface is the product; head gains are evidence it is useful where
capacity exists---not a claim that one classifier always wins.

\paragraph{Contributions.}
\begin{enumerate}[leftmargin=1.4em,itemsep=0.15em]
\item A trunk-agnostic personalization contract---wrap a frozen encoder,
  fit a Bayesian head on calibration embeddings, optionally apply an
  affine mid-tier, expose structured predictions---that runs unchanged
  across five classical trunks and one foundation trunk
  (\S\ref{sec:api}--\S\ref{sec:protocol}).
\item A multi-trunk map of where the stream head improves utility under
  severe shift, with subject-level confidence intervals (\S\ref{sec:head}).
\item A transferability analysis of the ladder across 18
  trunk\(\times\)dataset cells, distinguishing a reliable
  calibration-only-when-clean default (\(12/18\)) from a conditional
  strict-ordinal schedule (\(5/18\)) (\S\ref{sec:transfer}).
\item An accuracy--cost comparison positioning the head as a cheap default
  versus warm-start fine-tune and PEFT on every classical trunk and on
  foundation REVE (\S\ref{sec:cost}).
\item A head-family study (LDA/QDA/Softmax) that scopes QDA-dominance to a
  channel-mix-on-BNCI phenomenon rather than a universal head property,
  via a cross-stress replication (\S\ref{sec:heads}).
\end{enumerate}

\section{Personalizer API}
\label{sec:api}

\subsection{Contract}
\label{sec:contract}

\begin{figure}[t]
\centering
\includegraphics[width=0.92\linewidth]{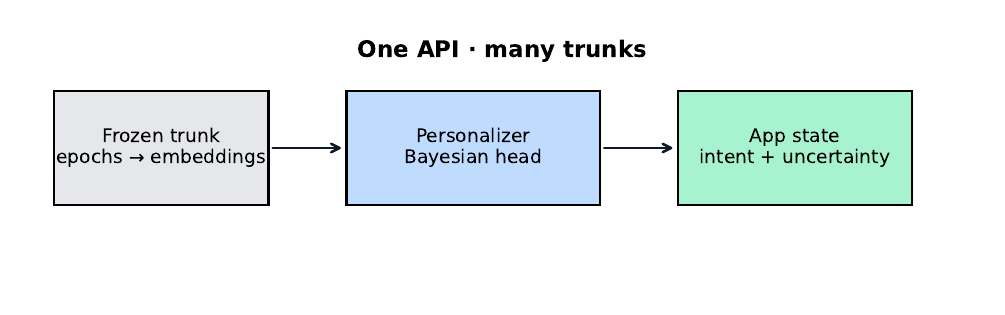}
\caption{Nimbus Personalizer surface: any frozen trunk that maps trials to
embeddings plugs into the same Bayesian head and structured app-state
contract (intent + uncertainty).}
\label{fig:api}
\end{figure}

The product chain is deliberately thin:

\begin{lstlisting}[language=Python,caption={Minimal Personalizer usage (product sketch).},label={lst:api}]
from nimbus_bci import wrap, Personalizer

enc = wrap(model.encode, model_id="trunk")
adapter = Personalizer(encoder=enc, head="lda",
                       classes=["left", "right"])
adapter.fit(X_cal, y_cal)
# Each state is a structured object, not a bare label:
# {"intent": "left", "confidence": 0.87, "uncertainty": 0.13,
#  "alternatives": [("right", 0.13)]}
states = adapter.predict(X_stream)
\end{lstlisting}

\paragraph{Encoder boundary.}
Integrators supply a frozen mapping from trials to embedding rows
\(Z\in\mathbb{R}^{n\times d}\) (callable, sklearn transformer, or torch
module method).
Adapters exist for BrainDecode-style trunks (including foundation
encoders such as REVE); the classical evaluation uses five
conv--attention trunks trained per subject then frozen.

\paragraph{Heads.}
Default head is LDA; QDA and an optional Softmax head share the same
Personalizer interface (Section~\ref{sec:heads}).
Multi-trunk transfer results below lock LDA for comparability with the
companion cost story.

\paragraph{App state.}
Predictions are not bare labels: each call returns intent hypotheses with
predictive uncertainty suitable for downstream decision presets.
Decision heuristics that choose \emph{whether} to escalate are out of
scope for this paper (companion control layer).

\subsection{Ladder levels under one API}
\label{sec:ladder}

We evaluate three personalization levels that the API can express without
changing the trunk:

\begin{itemize}[leftmargin=1.4em,itemsep=0.15em]
\item \textbf{L0 --- calibration-only Personalizer:} Bayesian head fit on
calibration embeddings only; no adaptation labels (adaptation-fit cost
counted as zero in the shared utility below).
\item \textbf{L1 --- stream Personalizer:} L0 head plus online updates on
labeled adaptation trials (LDA default).
\item \textbf{L2 --- affine map + head:} supervised affine map on
embeddings, then the same Personalizer head.
\end{itemize}
\paragraph{Notation.}
\emph{Factory} / \emph{trunk softmax} is the pretrained encoder's own
classifier (called L0 in the companion control-layer
paper~\citep{musienko2026ladder}).
Here L0 is always the calibration-only Personalizer head.
\emph{Frozen} always means encoder weights fixed; it never means ``no
Personalizer.'' Cost bake-offs report trunk softmax as a separate factory
baseline (Section~\ref{sec:ft}).

Full fine-tune (L3) is an escape hatch on every classical trunk in the
transfer suite (Sections~\ref{sec:wall},~\ref{sec:ft}); the EEGNet denser
baselines and Hub REVE bake-off (Section~\ref{sec:reve-ft}) add
intermediate adaptation arms (LoRA, last-layer FT) on one classical and one
foundation trunk.

\section{Evaluation protocol}
\label{sec:protocol}

\subsection{End-to-end pipeline}
\label{sec:pipeline}

\begin{figure}[H]
\centering
\includegraphics[width=0.98\linewidth]{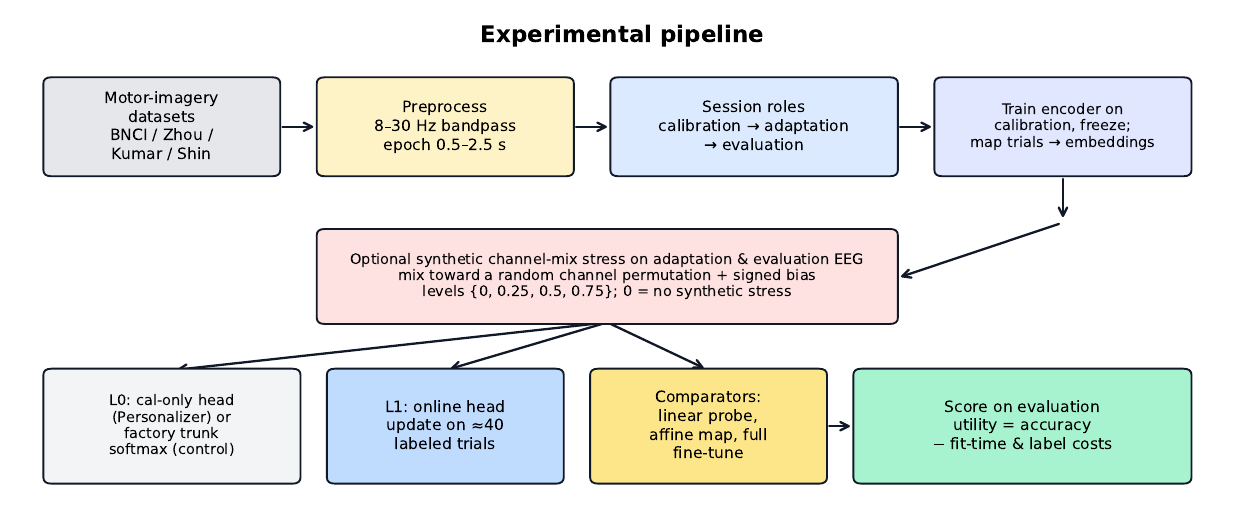}
\caption{Experimental data flow shared with the companion ladder paper.
Motor-imagery epochs are bandpass-filtered, assigned to calibration /
adaptation / evaluation sessions, used to train then freeze an encoder,
optionally stressed with synthetic channel mixing on adaptation and
evaluation EEG, mapped to embeddings, then personalized (L0/L1; optional
L2 affine map; L3 fine-tune hatch) and scored under
shared utility~\eqref{eq:U}.}
\label{fig:pipeline}
\end{figure}

Figure~\ref{fig:pipeline} is the single picture of what happens to the
data before any accuracy or wall number in this paper.

\paragraph{Datasets and sessions.}
We use public motor-imagery corpora via MOABB~\citep{moabb2021}
(Table~\ref{tab:datasets}).
Every subject uses a three-session geometry: \textbf{calibration} = first
session, \textbf{adaptation} = middle session(s),
\textbf{evaluation} = last session.
Labeled adaptation budget is \(n{=}40\) trials when available (Shin clamps to
\(n{=}20\) because the adaptation session is short).

\begin{table}[H]
\centering
\caption{Datasets in the 18-cell classical map (plus REVE separately).}
\label{tab:datasets}
\small
\begin{tabular}{lllp{5.2cm}}
\toprule
Dataset & Subjects & Classes & Notes \\
\midrule
BNCI2014-004 & 9 & 2 & Primary offline MI \\
Zhou2016 & 4 & 3 & Supporting; small \(n\) \\
Kumar2024 & 18 & 2 & Post-first sessions are online with feedback \\
Shin2017-A & 9 & 2 & Short adaptation session \(\Rightarrow\) \(n{=}20\) clamp \\
\bottomrule
\end{tabular}
\end{table}

\paragraph{Preprocessing.}
Epochs use a shared motor-imagery filter bank: bandpass
\textbf{8--30\,Hz}, time window \([0.5, 2.5]\)\,s after cue
(same band for all classical trunks so cross-trunk comparisons stay fair).
Foundation REVE cells that need a wider band use \textbf{1--40\,Hz} under
the same window; we do not mix bands inside a classical cell.

\paragraph{Trunk training and freeze.}
For each subject and trunk, the encoder is trained on \emph{clean}
calibration trials only, then frozen.
All personalization operates on embeddings \(Z\) obtained by a forward
pass through the frozen encoder; the Personalizer never updates trunk
weights except in the L3 escape hatch.

\paragraph{Trunks.}
Classical map: \{EEGNet, Shallow, Deep, Conformer, ATCNet\}.
Kumar runs the three trunks available in that suite
(EEGNet, Shallow, Conformer)---18 cells total.
Foundation REVE is evaluated separately under the same Personalizer
contract (Section~\ref{sec:reve-ft}).

\paragraph{Synthetic channel-mix stress.}
After the trunk is frozen we optionally perturb the concatenated
adaptation and evaluation \emph{raw multi-channel EEG} (not the
embeddings) with a controlled operator: a convex mix toward a random
channel permutation plus a signed per-channel bias, at stress levels
\(\{0,0.25,0.5,0.75\}\).
Level \(0\) disables the synthetic operator; natural session structure
may still differ from calibration (especially Kumar).
Calibration stays clean.
We do \emph{not} claim this exhausts physical electrode displacement,
SNR drop, or every clinical shift---those are out of scope here.

\paragraph{Adaptation arms scored on evaluation.}
L0 = calibration-only Personalizer; L1 = online head updates on the
labeled adaptation stream; L2 = optional affine map on embeddings then
head (protocol mid-tier); L3 = warm-start full fine-tune when the trunk
supports it.
A matched linear probe appears in cost figures as a control on the same
frozen embeddings, not as a ladder level.
Primary reported cells use \(n{=}40\) labeled adaptation trials
(Shin: \(n{=}20\)).

\paragraph{Utility.}
Levels are compared under a shared utility that trades classification
accuracy against adaptation fit time and labeled budget, defined with the
companion control-layer paper~\citep{musienko2026ladder} as
\begin{equation}
U = \mathrm{acc}
  - \lambda_w \cdot t_{\mathrm{adapt}}
  - \mu_n \cdot (n_{\mathrm{adapt}} / N_{\max}),
\label{eq:U}
\end{equation}
where \(t_{\mathrm{adapt}}\) is adaptation fit time in seconds (not
end-to-end latency) and \(n_{\mathrm{adapt}}\) is the number of labeled
stream trials spent (\(n_{\mathrm{adapt}}{=}0\) at L0). The two penalties
are in accuracy-equivalent units: \(\lambda_w\) is the accuracy cost of
one second of fit time, and \(\mu_n\) is the accuracy cost of a full
\(N_{\max}\)-trial labeled block. Our primary cells use
\(N_{\max}{=}80\) with \(\lambda_w{=}0.02\) and \(\mu_n{=}0.05\); in
product terms this reads as ``one second of adaptation fit time is priced
at \(2\)~pp of accuracy, and a full 80-trial labeled block at \(5\)~pp,''
which makes L0/L1/L2 comparable across trunks while keeping the expensive
option (full fine-tune) disfavored unless it buys substantial accuracy.
These specific weights are an engineering prior, not a fitted optimum; we
verified the central cost conclusion (capping at L1/L2 is preferred to
always fine-tuning) is robust across a \(5{\times}5\) grid
\(\lambda_w\in\{0.005,0.01,0.02,0.05,0.1\}\times\mu_n\in\{0.01,0.03,0.05,0.1,0.2\}\):
the fraction of cells where full fine-tune is the oracle utility winner
stays in \(0.00\)--\(0.28\) (BNCI), \(0.03\)--\(0.39\) (Kumar), and
\(0.00\)--\(0.50\) (Zhou), and the L1/L2 cap beats always-fine-tune on
mean utility at the fixed weights on all three datasets. Subject means;
exploratory.

\section{Where the Personalizer helps}
\label{sec:head}

\begin{figure}[H]
\centering
\includegraphics[width=0.98\linewidth]{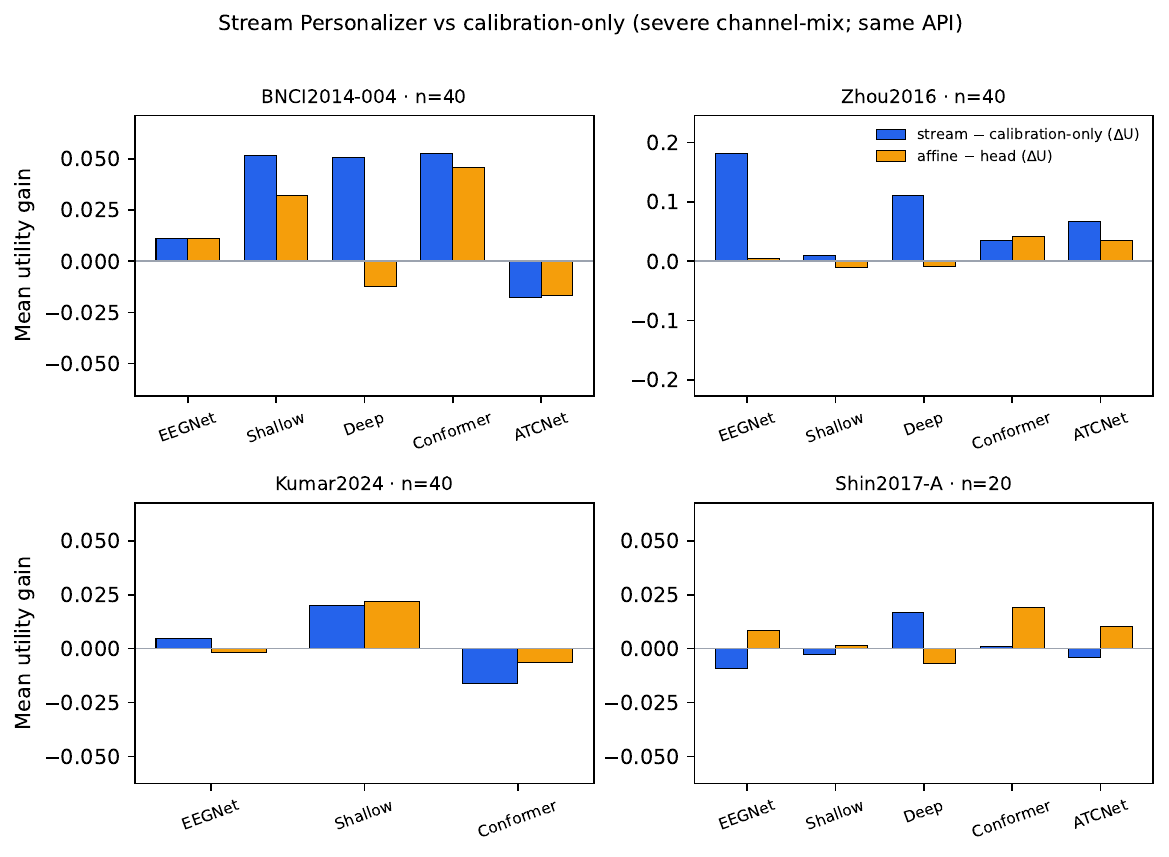}
\caption{Severe channel-mix stress: stream Personalizer vs cal-only (blue) is
the claim of this section; orange is optional affine mid-tier on top of
the same head.
Zhou: head helps on every trunk.
BNCI: strong on several trunks; ATCNet stays near cal-only.
Kumar/Shin: mixed---boundaries on the claim, not failures of the API.}
\label{fig:head}
\end{figure}

\begin{figure}[H]
\centering
\includegraphics[width=0.98\linewidth]{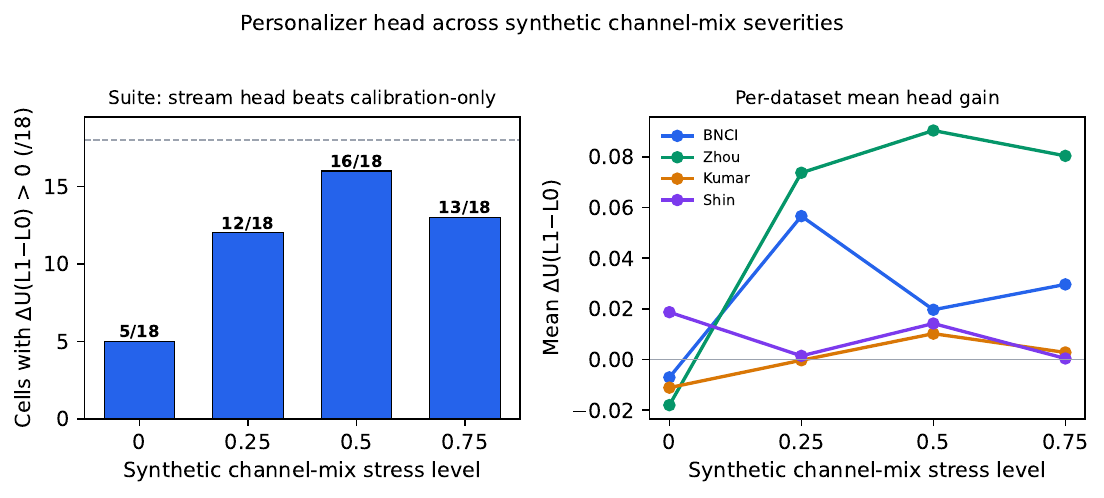}
\caption{Personalizer head across severities (18 trunk\(\times\)dataset
cells; same channel-mix protocol).
Left: cells with positive mean \(\Delta U_{\mathrm{L1}-\mathrm{L0}}\).
Right: per-dataset mean of that gain.
Mid shift already recovers; severe is the detailed slice above, not the
only supporting point.}
\label{fig:sev-sweep}
\end{figure}

The claim in this section concerns stream Personalizer (L1): where does
it beat cal-only Personalizer (L0)?
We measure the full synthetic channel-mix stress grid
\(\{0,0.25,0.5,0.75\}\) (Section~\ref{sec:protocol}); trunk-level detail
below focuses on severity \(0.75\), with the mid-severity sweep in
Figure~\ref{fig:sev-sweep}.
Figure~\ref{fig:head} plots severe gains
(\(\Delta U_{\mathrm{L1}-\mathrm{L0}}\), blue) on every trunk\(\times\)dataset
cell; orange bars are an \emph{optional} mid-tier step in the same API
(affine-\(Z\) then head, \(\Delta U_{\mathrm{L2}-\mathrm{L1}}\)), not a
separate component.
Figure~\ref{fig:head} and Table~\ref{tab:severe} report which ladder
level wins by mean utility at severity \(0.75\).
L1 and L2 are both cheap relative to full fine-tune (wall times within
\(\sim\)1.2--1.8\(\times\) of each other); the priced contrast with L3 is
deferred to Section~\ref{sec:cost}.

\paragraph{Zhou --- clearest win.}
Under severe shift, \(\Delta U_{\mathrm{L1}-\mathrm{L0}}>0\) for every
trunk (range \(\approx +0.009\) to \(+0.18\)): one API, five trunks, head
recovery when capacity exists.
Affine is not automatic: it adds utility on EEGNet, Conformer, and
ATCNet, while Shallow and Deep remain L1-best by mean \(U\)
(Table~\ref{tab:severe}).

\paragraph{BNCI --- often useful, not automatic.}
Head gains are positive for Shallow, Conformer, Deep, and EEGNet
(\(\approx +0.01\) to \(+0.05\)); ATCNet's best mean utility remains L0
in this cell.
Deep stays at L1; Shallow/EEGNet/Conformer take the optional L2 step.
The API still runs; recovery depends on trunk capacity.

\paragraph{Kumar / Shin --- boundaries.}
On Kumar severe the winner splits across the three trunks: stream head
(L1) on EEGNet, affine (L2) on Shallow, and cal-only (L0) on Conformer.
Shin (short adaptation session, \(n{=}20\)) is likewise mixed: cal-only
(L0) on EEGNet/Shallow, stream head (L1) on Deep, affine (L2) on
Conformer/ATCNet.
These cells bound overclaim: Personalizer is the default
\emph{integration} path, not a guarantee that L1 beats L0 under every
shift.

\begin{table}[!htbp]
\centering
\caption{Severe cell (\(0.75\)): mean-utility winner under the shared
Personalizer ladder. Shin uses \(n{=}20\); others \(n{=}40\).}
\label{tab:severe}
\small
\begin{tabular}{llccccc}
\toprule
Dataset & \(n\) & EEGNet & Shallow & Deep & Conformer & ATCNet \\
\midrule
BNCI & 9 & L2 & L2 & L1 & L2 & L0 \\
Zhou & 4 & L2 & L1 & L1 & L2 & L2 \\
Kumar & 18 & L1 & L2 & --- & L0 & --- \\
Shin & 9 & L0 & L0 & L1 & L2 & L2 \\
\bottomrule
\end{tabular}
\end{table}

\paragraph{Not only severe.}
Figure~\ref{fig:sev-sweep} aggregates the same 18 cells across the full
severity grid.
At severity \(0\), stream head beats cal-only in only \(5/18\) cells
(cal-only-when-clean).
Already at \(0.25\) the count rises to \(12/18\), peaks at \(16/18\) for
\(0.5\), and remains \(13/18\) at the severe slice \(0.75\).
Per-dataset mean \(\Delta U_{\mathrm{L1}-\mathrm{L0}}\) turns positive
under mid shift on Zhou/BNCI; Kumar stays near zero---a boundary, not a
severe-only artifact.
This sweep uses one shift family (synthetic channel mixing); other
perturbations (electrode drop, SNR) are out of scope here.

\FloatBarrier
\section{Transferability of the ladder across cells}
\label{sec:transfer}

\begin{figure}[t]
\centering
\includegraphics[width=\linewidth]{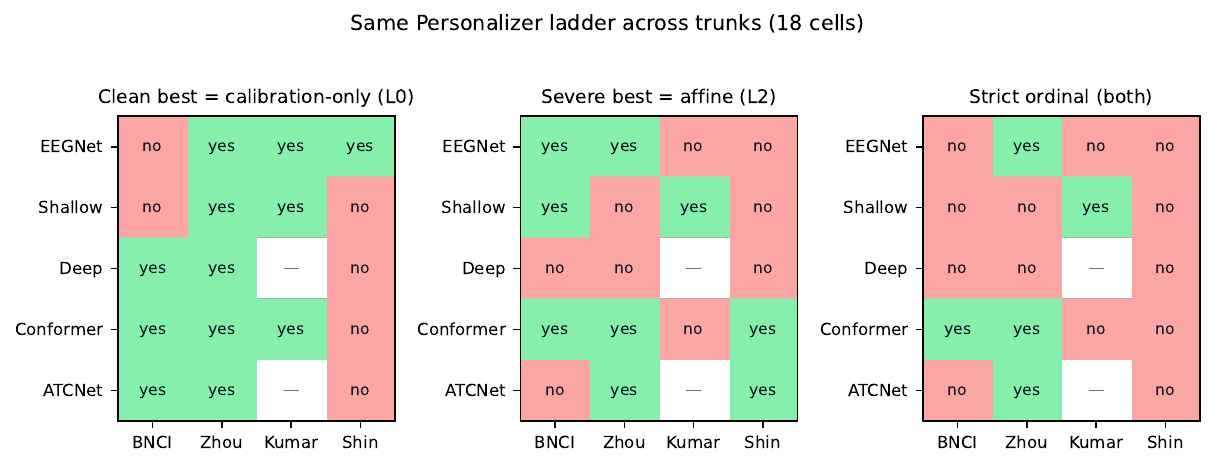}
\caption{Transfer map for the same Personalizer ladder (18 cells).
Green = criterion holds; red = fails; dash = trunk not in suite.
Cal-only-when-clean transfers often; requiring severe\(\to\)L2 is a
strict mid-tier schedule and is much more conditional.}
\label{fig:map}
\end{figure}

Having shown where the stream head helps, we ask what \emph{strict}
ordinal pattern repeats across the 18 trunk\(\times\)dataset cells
(Figure~\ref{fig:map}, Table~\ref{tab:inv}).
This is a secondary boundary analysis---not the main contribution.

\begin{itemize}[leftmargin=1.4em,itemsep=0.15em]
\item \textbf{Clean \(\to\) L0:} cal-only Personalizer best by mean \(U\)
at severity \(0\).
\item \textbf{Severe \(\to\) L2:} optional affine mid-tier best at
severity \(0.75\) (stronger than ``L1 helps''; Section~\ref{sec:head}).
\item \textbf{Strict ordinal:} both of the above (suite flag
\texttt{principle\_holds}).
\end{itemize}

\begin{table}[!htbp]
\centering
\caption{Pooled and per-dataset transfer counts (exploratory; mean-\(U\)
winners). Strict ordinal \(=\) clean\(\to\)L0 and severe\(\to\)L2.}
\label{tab:inv}
\small
\begin{tabular}{lccc}
\toprule
Slice & Clean\(\to\)L0 & Severe\(\to\)L2 & Strict ordinal \\
\midrule
Zhou & \(5/5\) & \(3/5\) & \(3/5\) \\
BNCI & \(3/5\) & \(3/5\) & \(1/5\) \\
Kumar & \(3/3\) & \(1/3\) & \(1/3\) \\
Shin & \(1/5\) & \(2/5\) & \(0/5\) \\
\midrule
\textbf{Pooled} & \(\mathbf{12/18}\) & \(9/18\) & \(\mathbf{5/18}\) \\
\bottomrule
\end{tabular}
\end{table}

\paragraph{Interpretation.}
The Personalizer \emph{API} is universal in the engineering sense: one
contract, many trunks, runnable ladder levels.
Cal-only-when-clean is the reliable operational default
(\(12/18\); BNCI EEGNet/Shallow are clean exceptions after recalculation).
Requiring severe\(\to\)L2 is a \emph{strict} mid-tier schedule: it holds
on \(9/18\) cells (Zhou \(3/5\)), and the joint ordinal flag on \(5/18\).
That is consistent with Section~\ref{sec:head}: stream head recovery is
broader than ``always take affine.''
Policy escalation remains a design principle for recoverable structure,
not a universal law---aligned with the companion
paper~\citep{musienko2026ladder}, now at trunk resolution.

\paragraph{Statistical caveat (counts are small).}
The pooled counts rest on 18 trunk\(\times\)dataset cells.
A Wilson 95\% CI on the \(12/18\) cal-only-when-clean rate is
\([0.44,0.84]\): the interval includes the \(0.5\) chance rate, and against
that null the one-sided \(p=0.119\) does not survive Holm/BH correction
across the count-claim family.
The strict-ordinal \(5/18\) (\([0.12,0.51]\)) is not significant against the
\(0.25\) null for two independent conditions (\(p=0.481\)), and severe\(\to\)L2
(\(9/18\), \([0.29,0.71]\)) is at chance.
The counts therefore support a \emph{soft, dataset-dependent} tendency (strong
on Zhou, weak on Shin) rather than a quantified universal rule; ``reliable
default'' above means operationally, not statistically at \(n{=}18\).

\FloatBarrier
\section{Adaptation cost}
\label{sec:cost}

Cost evidence has one \emph{core} comparison on every classical trunk
(Personalizer vs warm-start full FT, plus a linear-probe control) and two
\emph{spotlight} extended baseline configurations (EEGNet classical
reference; Hub REVE foundation reference).
Last-layer FT and LoRA---intermediate adaptation methods between the head
and full fine-tune---stay in the spotlights: they are architecture-specific
rather than universally available across the multi-trunk map.

\subsection{Adaptation fit cost across trunks}
\label{sec:wall}

\begin{figure}[!htbp]
\centering
\includegraphics[width=\linewidth]{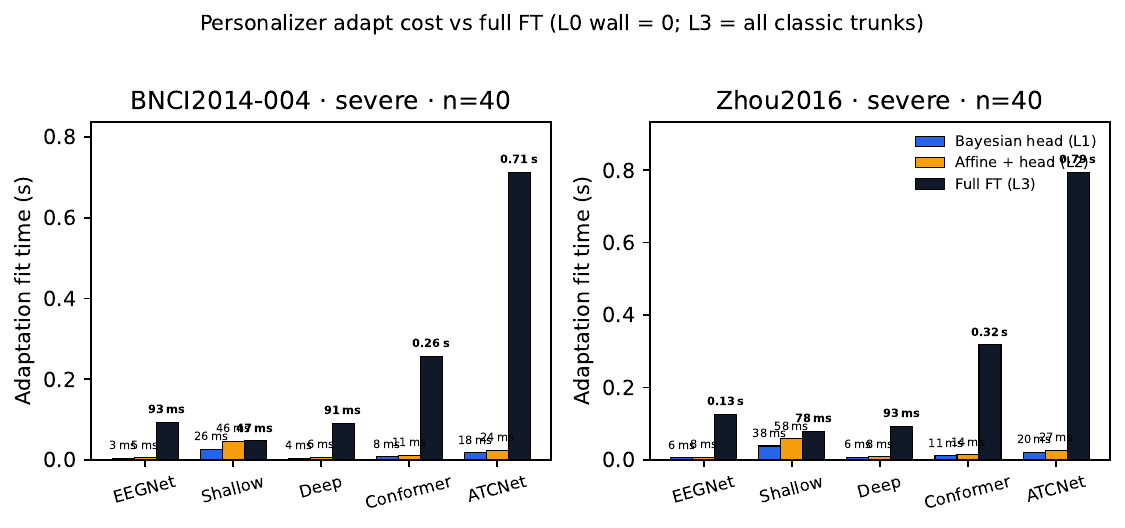}
\caption{Adaptation fit time under the same Personalizer API (severe,
\(n{=}40\); CUDA).
Blue: stream Personalizer (L1); orange: affine$+$head (L2); black: full FT
(L3) on every classical trunk.
Independent $y$-scales per dataset.
Cal-only L0 stream wall is zero.}
\label{fig:wall}
\end{figure}

Figure~\ref{fig:wall} reports mean adaptation \emph{fit} wall time at
severity \(0.75\) (same transfer protocol; subject means): L1/L2/L3 on
every classical trunk.
Cal-only L0 has zero \emph{stream} adaptation fit by construction.
On CUDA, L1/L2 stay in the millisecond-to-tens-of-ms band on BNCI and
sub-second on Zhou, while L3 is \(\sim\)0.05--0.8\,s depending on trunk
(ATCNet the slowest)---still far above the Personalizer head.
Accuracy for the same L3 cells is in Section~\ref{sec:ft}.
This is fit cost, not inference latency.

\subsection{Personalizer vs full fine-tune (all classical trunks)}
\label{sec:ft}

\begin{figure}[H]
\centering
\includegraphics[width=0.98\linewidth]{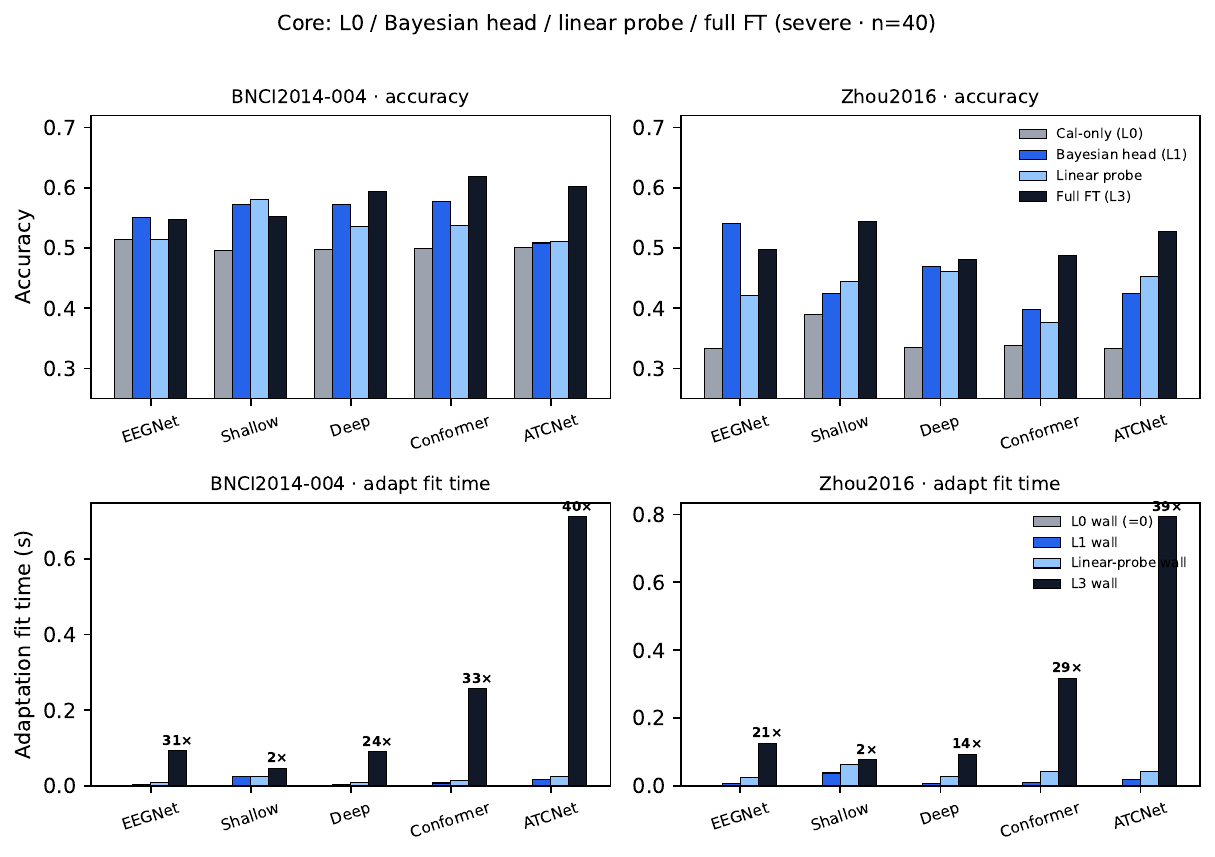}
\caption{Core multi-trunk cost block (severe, \(n{=}40\)).
Top: accuracy for cal-only (L0), stream Personalizer (L1), linear probe
on frozen \(Z\), and warm-start full FT (L3).
Bottom: adapt fit wall (L0 stream wall is zero) with L3/L1 ratios.
Per-trunk recovery ratios are in Table~\ref{tab:ft-trunks}.}
\label{fig:ft-trunks}
\end{figure}

\textbf{Core block.}
On every classical trunk we compare cal-only Personalizer (L0), stream
Personalizer (L1), a matched sklearn linear probe on the same frozen
\(Z\), and warm-start full FT (L3) under the transfer protocol
(Figure~\ref{fig:ft-trunks}, Table~\ref{tab:ft-trunks}; severe,
\(n{=}40\), CUDA).
Wall times alone were in Figure~\ref{fig:wall}; here we close the
accuracy--cost loop and ask whether Bayesian personalization is just
``any head on \(Z\).''

\begin{table}[H]
\centering
\caption{Core severe \(n{=}40\): mean accuracy for cal-only L0, Bayesian
L1, linear probe (LP), and warm-start L3; L3/L1 wall ratio.
Recovery \(=(\mathrm{acc}_{\mathrm{L1}}-\mathrm{acc}_{\mathrm{L0}})/
(\mathrm{acc}_{\mathrm{L3}}-\mathrm{acc}_{\mathrm{L0}})\).
``$\geq$FT'' means \(\mathrm{acc}_{\mathrm{L1}}\ge\mathrm{acc}_{\mathrm{L3}}\).}
\label{tab:ft-trunks}
\small
\begin{tabular}{llcccccc}
\toprule
Dataset & Trunk & Acc L0 & Acc L1 & Acc LP & Acc L3 & Recovery & Wall L3/L1 \\
\midrule
BNCI & EEGNet & 0.514 & 0.550 & 0.514 & 0.548 & \(\geq\)FT & \({\sim}31\times\) \\
BNCI & Shallow & 0.495 & 0.572 & 0.580 & 0.553 & \(\geq\)FT & \({\sim}2\times\) \\
BNCI & Deep & 0.497 & 0.573 & 0.536 & 0.594 & \(78\%\) & \({\sim}24\times\) \\
BNCI & Conformer & 0.500 & 0.578 & 0.537 & 0.618 & \(66\%\) & \({\sim}33\times\) \\
BNCI & ATCNet & 0.501 & 0.508 & 0.510 & 0.602 & \(8\%\) & \({\sim}40\times\) \\
Zhou & EEGNet & 0.333 & 0.540 & 0.422 & 0.498 & \(\geq\)FT & \({\sim}21\times\) \\
Zhou & Shallow & 0.390 & 0.425 & 0.445 & 0.543 & \(23\%\) & \({\sim}2\times\) \\
Zhou & Deep & 0.335 & 0.470 & 0.462 & 0.482 & \(92\%\) & \({\sim}14\times\) \\
Zhou & Conformer & 0.338 & 0.398 & 0.377 & 0.488 & \(40\%\) & \({\sim}29\times\) \\
Zhou & ATCNet & 0.333 & 0.425 & 0.453 & 0.527 & \(47\%\) & \({\sim}39\times\) \\
\bottomrule
\end{tabular}
\end{table}

\paragraph{Accuracy.}
On most trunks the Bayesian head recovers a large fraction of the L3 gain
over cal-only (often \({\sim}65\%\)--\(90\%\)) and beats the matched linear
probe (EEGNet/Deep/Conformer on both datasets).
Shallow and ATCNet are the probe-competitive cells: LP can edge L1 by a
few points while both stay far below L3 wall cost---so Personalizer is
not magically always best on raw accuracy, but it remains the chosen
default (same API, UQ/\texttt{BrainState}).
BNCI Shallow is still the Pareto win vs FT: L1 \emph{matches or beats} L3
at \({\sim}2\times\) wall.
ATCNet on BNCI remains the capacity boundary where L3 buys accuracy the
cheap heads do not.

\paragraph{Mid severity, not only severe.}
Figure~\ref{fig:ft-trunks} and Table~\ref{tab:ft-trunks} use severity
\(0.75\) as the hard accuracy slice (largest FT residual).
At mid severity the head is closer to FT: across BNCI+Zhou subject cells,
stream L1 matches or beats warm-start L3 on accuracy in about half of
cells at \(0.25\)/\(0.5\) (\({\sim}52\%\)/\({\sim}57\%\)) versus
\({\sim}43\%\) at \(0.75\).
Wall L3/L1 ratios stay in the same expensive band at every severity, so
mid shift strengthens the ``head first'' default; severe is where FT most
often still buys residual pp.

\paragraph{Cost.}
Except Shallow (\({\sim}2\times\)), L3/L1 wall ratios sit in the
\({\sim}14\times\)--\({\sim}41\times\) band on CUDA---the same qualitative
pricing as the REVE bake-off: FT is the expensive hatch.
Under~\eqref{eq:U}, stopping at L1/L2 is usually preferred unless the
accuracy budget explicitly buys L3.

\begin{figure}[H]
\centering
\includegraphics[width=0.95\linewidth]{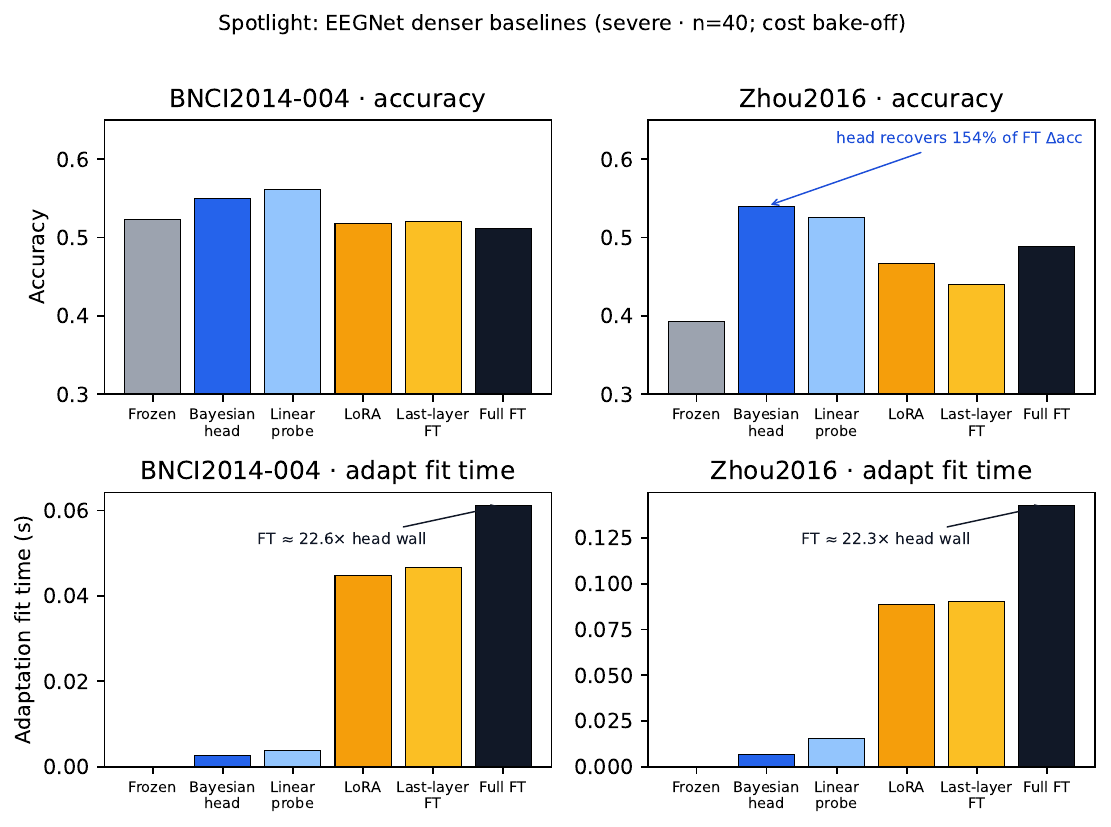}
\caption{Spotlight --- EEGNet denser baselines (cost bake-off; severe,
\(n{=}40\)): trunk-softmax frozen, Bayesian head, linear probe, LoRA,
last-layer FT, full FT.
Same visual grammar as the REVE spotlight (head $\to$ linear $\to$
last/full); not a claim that every classical trunk exposes the same
intermediate adaptation knobs.}
\label{fig:ft}
\end{figure}

\paragraph{Spotlight: EEGNet denser baselines.}
For one classical reference trunk we keep a denser baseline configuration
against trunk-softmax frozen, LoRA, and last-layer FT
(Figure~\ref{fig:ft}, Table~\ref{tab:ft}).
This is a spotlight, not the multi-trunk claim: last-layer / LoRA are
EEGNet-specific intermediate adaptation methods.
Bake-off walls are protocol-different from ladder L1 (hence
\({\sim}23\times\) on the EEGNet bake-off vs \({\sim}21\times\)--\(40\times\) in
Table~\ref{tab:ft-trunks}); both say FT is not the default.

\begin{table}[H]
\centering
\caption{EEGNet severe \(n{=}40\) cost bake-off means (trunk-softmax
frozen baseline; final evaluation metrics).
When FT \(\le\) frozen we report absolute accuracies (recovery ratio vs FT
\(\Delta\mathrm{acc}\) is not meaningful).
Bake-off wall ratio \(=\mathrm{wall}_{\mathrm{FT}}/\mathrm{wall}_{\mathrm{head}}\).}
\label{tab:ft}
\small
\begin{tabular}{lccccc}
\toprule
Dataset & Acc frozen & Acc head & Acc FT & vs FT & Bake-off wall \\
\midrule
BNCI & 0.523 & 0.550 & 0.512 & beats FT & \({\sim}23\times\) \\
Zhou & 0.393 & 0.540 & 0.488 & beats FT & \({\sim}22\times\) \\
\bottomrule
\end{tabular}
\end{table}

\FloatBarrier
\subsection{Spotlight: foundation REVE}
\label{sec:reve-ft}

\begin{figure}[H]
\centering
\includegraphics[width=0.95\linewidth]{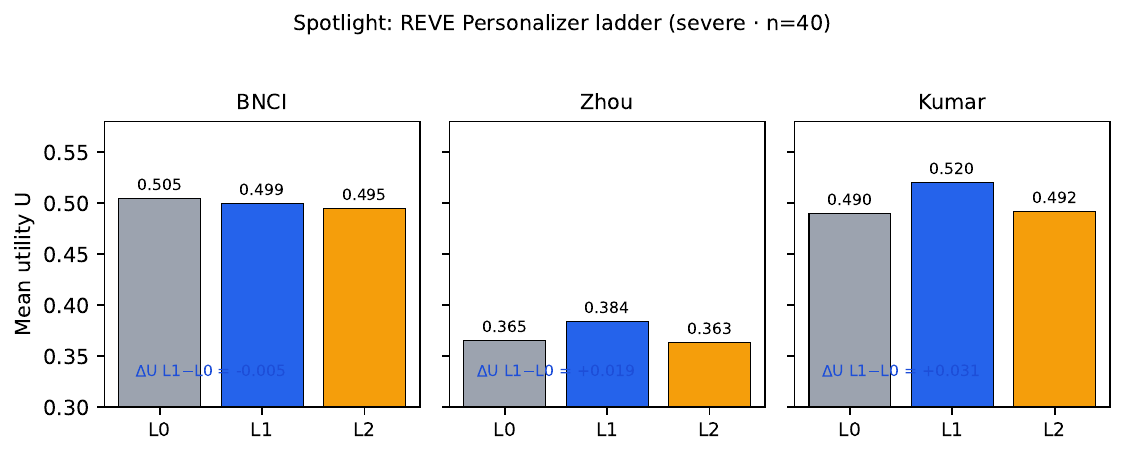}
\caption{Frozen REVE under the same L0/L1/L2 Personalizer ladder (severe,
\(n{=}40\)).
Head (L1) helps on Zhou and Kumar; BNCI remains a capacity boundary.}
\label{fig:reve-ladder}
\end{figure}

\begin{figure}[H]
\centering
\includegraphics[width=0.95\linewidth]{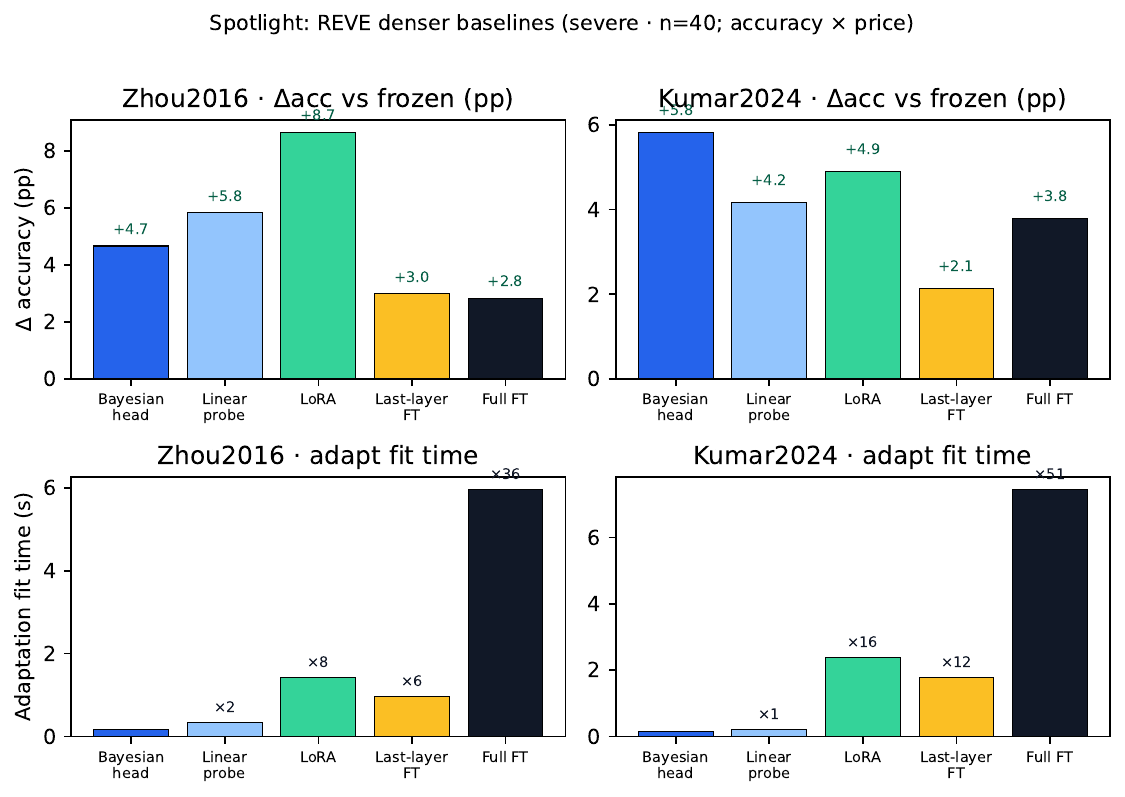}
\caption{Spotlight --- Hub REVE denser baselines (severe, \(n{=}40\)).
Top: \(\Delta\mathrm{acc}\) vs frozen Personalizer (cal fit only).
Bottom: adaptation fit wall.
Arms: head / linear / LoRA / last / full --- LoRA is the honest PEFT
intermediate method on a large pretrained trunk (EEGNet LoRA alone is not).
Last-layer / full FT use the same mean-pooled Linear head as LoRA
(not the encoder's raw token-level classifier); full FT uses differential LR.}
\label{fig:reve-ft}
\end{figure}

The same Personalizer ladder runs on Hub
REVE~\citep{elouahidi2025reve} without a new API
(Figure~\ref{fig:reve-ladder}): under severe shift, L1 lifts mean utility
on Zhou and Kumar; BNCI (3-channel) stays near chance---capacity, not
contract, is the limit.
As a foundation spotlight we then price the head against linear probe,
point LoRA (FFN/head linears; the PEFT intermediate method that is unfair
to judge on tiny EEGNet alone), and mean-pooled last-layer / full FT under a
matched label budget (Figure~\ref{fig:reve-ft}, Table~\ref{tab:reve-ft};
Modal L40S, warm-start CE with early stopping).
Full FT uses a pooled Linear head with differential LR (backbone
\(10^{-5}\), head \(10^{-3}\)) rather than the encoder's raw token-level
classifier, which is a weak strawman on multi-channel REVE.

\begin{table}[H]
\centering
\caption{Frozen REVE severe \(n{=}40\): Personalizer vs LoRA / last-layer /
full FT (mean accuracy and adapt wall; Modal L40S; pooled-head FT recipe).
\(\Delta\mathrm{acc}\) is vs frozen Personalizer (cal fit only).
Wall ratio \(=\mathrm{wall}/\mathrm{wall}_{\mathrm{head}}\).
Zhou: 4; Kumar: 18.}
\label{tab:reve-ft}
\small
\begin{tabular}{lccccccc}
\toprule
Dataset & Acc frozen & Acc head & Acc LoRA & Acc last & Acc full
  & Wall LoRA/head & Wall full/head \\
\midrule
Zhou & 0.365 & 0.412 & 0.452 & 0.395 & 0.393 & \({\sim}8\times\) & \({\sim}36\times\) \\
Kumar & 0.490 & 0.548 & 0.539 & 0.511 & 0.528 & \({\sim}16\times\) & \({\sim}51\times\) \\
\bottomrule
\end{tabular}
\end{table}

\paragraph{Price.}
Personalizer adapt wall stays sub-second (\({\sim}0.15\,\mathrm{s}\));
point LoRA sits at \({\sim}8\times\)--\(16\times\) head wall; pooled
last-layer / full FT at \({\sim}6\times\)--\(51\times\).

\paragraph{Pareto.}
Even with a fair pooled-head FT recipe (differential LR; Dropout/BN in
eval), short full FT at \(n{=}40\) does \emph{not} overturn the head
default: on Kumar the head still leads (\(+5.8\,\mathrm{pp}\) vs LoRA
\(+4.9\), full \(+3.8\), last \(+2.1\)); on Zhou LoRA is the best PEFT
intermediate method (\(+8.7\,\mathrm{pp}\) at \({\sim}8\times\)), while pooled
last/full stay near \(+3\,\mathrm{pp}\) at much higher wall.
A prior last-layer arm using the encoder's raw token-level classifier looked
stronger but was an unfair high-capacity strawman relative to the encode
contract---we do not cite it.

\paragraph{Claim boundary.}
We do not claim that frozen REVE + Personalizer matches cal-trained EEGNet
absolute accuracy.
We claim a cheap API mid-point on a foundation trunk: escalate to LoRA /
encoder FT only when the accuracy budget justifies
\({\sim}8\times\)--\(50\times\) adapt cost.

\FloatBarrier
\section{Head choice: LDA versus QDA versus Softmax}
\label{sec:heads}

\begin{figure}[H]
\centering
\includegraphics[width=\linewidth]{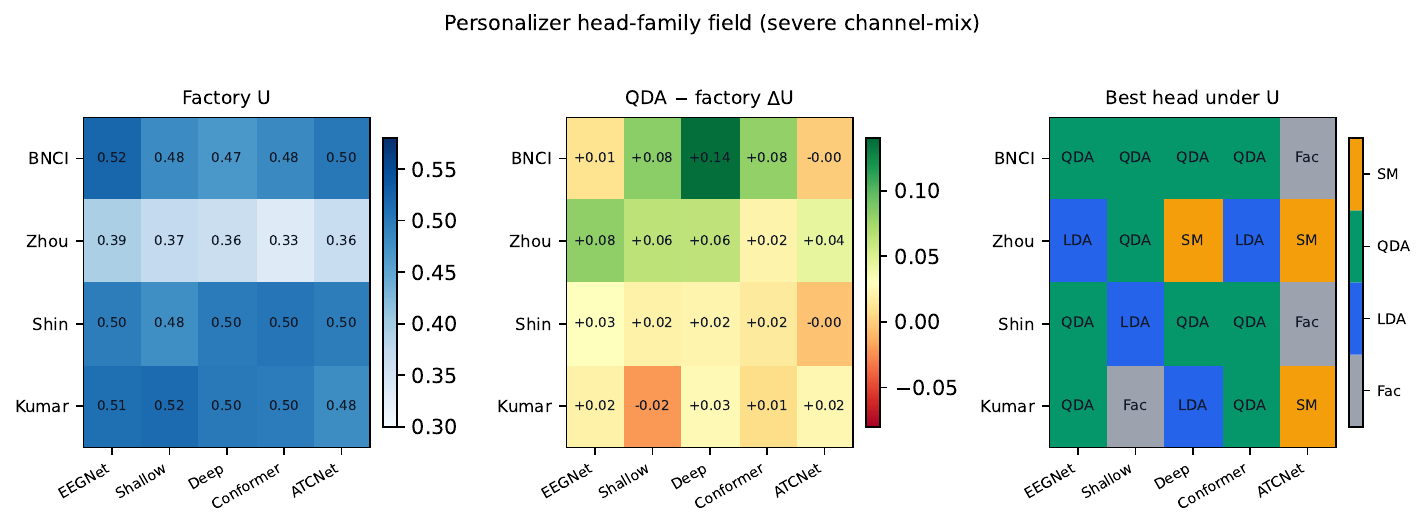}
\caption{Personalizer head-family field across 5 trunks \(\times\) 4 datasets
(severe channel-mix \(0.75\)). Left: factory (trunk softmax) utility---broken
everywhere under severe shift. Middle: QDA\(\!-\!\)factory \(\Delta U\)
(QDA helps where \(>0\)); QDA-dominance is concentrated on BNCI. Right: best head
under \(U\) per cell---no single head wins across the field.
Kumar Deep/ATCNet are extra head-bakeoff cells outside the 18-cell ladder map.}
\label{fig:head-field}
\end{figure}

The multi-trunk transfer map (Sections~\ref{sec:head}--\ref{sec:transfer})
locks LDA as a fixed baseline. This section asks an API-design question
that the baseline cannot answer: should the Personalizer contract
\emph{hardcode} one classifier, or expose an interchangeable head parameter
(\texttt{head="lda"|"qda"|"softmax"})? A single hardcoded head is only
safe if one classifier is universally best; if head optimality is
dataset- or stress-dependent, a modular head is the correct contract.
We therefore ran QDA and Softmax alongside LDA on every classical
trunk \(\times\) dataset cell (severe and clean, \(n{=}40\);
Table~\ref{tab:heads-multitrunk}).
On the EEGNet reference trunk the picture is: QDA edges LDA on BNCI
(\(+0.008\) \(U\), accuracy \(0.558\) vs \(0.550\)) but loses on Zhou
(\(-0.040\)), while Softmax is roughly two orders of magnitude slower to
fit (\({\sim}0.46\)s vs \({\sim}0.004\)s) without a utility payoff---so
across the field no single head dominates, and under the shared utility
QDA is the \emph{best} head on BNCI for every trunk while on Zhou the
best head is dataset-dependent.
A caveat matters for the API design: channel mixing---the operator used
throughout this paper---is the stress family where quadratic boundaries
happen to shine, so the BNCI QDA edge in Figure~\ref{fig:head-field} is
the favorable case. Appendix~\ref{app:cross-stress} validates the same
heads under three further stress families and finds LDA more stable
overall, which is precisely why LDA remains the API default rather than
QDA.

\begin{table}[H]
\centering
\caption{Multi-trunk head family under the shared utility, severe channel-mix
(\(0.75\); BNCI/Zhou/Kumar \(n{=}40\), Shin short-stream \(n{=}10\)). Mean
utility \(U\) by trunk \(\times\) dataset; \textbf{bold} = best head per row.
QDA is best Personalizer head in \(13/20\) cells; on BNCI it wins \(4/5\)
overall (ATCNet stays factory-best under \(U\)); elsewhere the winner is
dataset-dependent.}
\label{tab:heads-multitrunk}
\small
\begin{tabular}{llccccc}
\toprule
Dataset & Trunk & \(U\) frozen & \(U\) LDA & \(U\) QDA & \(U\) Softmax & best \\
\midrule
BNCI & eegnet    & 0.522 & 0.525 & \textbf{0.533} & 0.496 & QDA \\
BNCI & shallow   & 0.481 & 0.546 & \textbf{0.564} & 0.518 & QDA \\
BNCI & deep      & 0.468 & 0.548 & \textbf{0.604} & 0.509 & QDA \\
BNCI & conformer & 0.483 & 0.552 & \textbf{0.564} & 0.496 & QDA \\
BNCI & atcnet    & \textbf{0.501} & 0.483 & 0.500 & 0.481 & frozen \\
\midrule
Zhou & eegnet    & 0.393 & \textbf{0.515} & 0.475 & 0.507 & LDA \\
Zhou & shallow   & 0.372 & 0.399 & \textbf{0.436} & 0.361 & QDA \\
Zhou & deep      & 0.362 & 0.448 & 0.425 & \textbf{0.456} & SM \\
Zhou & conformer & 0.335 & \textbf{0.371} & 0.357 & 0.351 & LDA \\
Zhou & atcnet    & 0.365 & 0.398 & 0.408 & \textbf{0.413} & SM \\
\midrule
Shin & eegnet    & 0.497 & 0.511 & \textbf{0.526} & 0.485 & QDA \\
Shin & shallow   & 0.478 & \textbf{0.502} & 0.498 & 0.496 & LDA \\
Shin & deep      & 0.498 & 0.502 & \textbf{0.521} & 0.508 & QDA \\
Shin & conformer & 0.503 & 0.507 & \textbf{0.519} & 0.490 & QDA \\
Shin & atcnet    & \textbf{0.497} & 0.492 & 0.492 & 0.475 & frozen \\
\midrule
Kumar & eegnet    & 0.511 & 0.508 & \textbf{0.532} & 0.493 & QDA \\
Kumar & shallow   & \textbf{0.516} & 0.494 & 0.495 & 0.482 & frozen \\
Kumar & deep      & 0.501 & \textbf{0.540} & 0.527 & 0.499 & LDA \\
Kumar & conformer & 0.497 & 0.491 & \textbf{0.504} & 0.481 & QDA \\
Kumar & atcnet    & 0.480 & 0.494 & 0.504 & \textbf{0.505} & SM \\
\bottomrule
\end{tabular}
\end{table}

On BNCI severe, QDA is the top arm for \(4/5\) trunks (ATCNet stays factory-best
under \(U\)); the EEGNet QDA gain over LDA is modest (\(+0.008\) \(U\)).
Across the \(20\)-cell matrix QDA is the best Personalizer head in
\(13/20\) cells. QDA dominance does
\emph{not} generalize: on Zhou, Shin, and Kumar the best head is
dataset-dependent (LDA/QDA/Softmax/frozen each win on different cells), and
on Kumar/Shin the factory decoder still wins some cells under \(U\).
The honest reading is that head-choice under severe shift is
\textbf{not governed by a single rule}; it depends on the trunk\(\times\)dataset
pair, and no head (including QDA) is universally best.

\paragraph{How head choice varies.}
Head choice varies along three axes. \emph{Statistically}, subject-level
paired-bootstrap 95\% CIs first average repeated trunk measurements within
each participant, then resample participants. The QDA-over-LDA gain on BNCI
is \(+0.022\) \(U\), \([+0.001,+0.049]\), \(n{=}9\) subjects; the
L1-over-L0 gain on Zhou is \(+0.080\), \([+0.002,+0.162]\), \(n{=}4\).
For BNCI, Kumar, and Shin, the L1--L0 intervals span zero.
\emph{Across datasets}, on a held-out Shin/Kumar sweep the structural
claim (the default head closes most of the room available to denser
intermediate adaptation methods) \textbf{generalizes}, but the head-dominance pattern does
not: QDA beats factory under \(U\) in \(9/10\) BNCI/Zhou cells but only
\(7/10\) of the pooled field, and the BNCI/Zhou ``ATCNet capacity
boundary'' \textbf{inverts} on Shin (\(\Delta U{=}{+}0.043\)) and Kumar
(\(\Delta U{=}{+}0.042\)).
\emph{Across stress families}, re-running the head family under three
further synthetic operators (amplitude scaling, channel dropout, SNR
noise; 5 trunks \(\times\) 3 datasets) the pooled QDA-minus-LDA effect is
near zero or negative (\(-0.015\), \(-0.006\), \(+0.001\); only feature
shift trends positive at \(+0.007\)); the full per-cell sweep is reported
in Appendix~\ref{app:cross-stress}.
Together these results place QDA's advantage in the
\emph{channel-mix-on-BNCI} setting. LDA therefore remains the API default,
with QDA available to benchmark per trunk\(\times\)dataset\(\times\)stress.
Figure~\ref{fig:head-field} maps the full field; the 18-cell ladder map
uses Kumar\(\times\)\{EEGNet, Shallow, Conformer\}, while
Table~\ref{tab:heads-multitrunk} and Figure~\ref{fig:head-field} also
report Kumar Deep/ATCNet from the same harness as extra cells outside that
count.

\paragraph{Does QDA clear the label tax?}
At \(n{=}40\) the utility's label term is \(\mu_n(n/80)=0.025\). On BNCI QDA
clears it for four of five trunks (only ATCNet stays at parity with the factory
decoder). On Zhou every head clears the label tax over frozen, but the best head
varies. On Kumar, where the frozen decoder is comparatively strong under severe
shift, the label tax is \emph{not} cleared on Shallow (\(1/5\) cells)---spending
labels on a head loses utility versus doing nothing there. Label-tax clearance
is therefore common but not universal.

\paragraph{Default head.}
LDA remains the API default (no optional dependency, chosen for
comparability across the multi-trunk map).
The multi-trunk field downgrades QDA from a candidate universal default to a
\textbf{BNCI-concentrated strong option}: on BNCI it is the best head under
\(U\) for \(4/5\) trunks (ATCNet stays factory-best), but on the other three
datasets the integrator should benchmark LDA/QDA/Softmax per
trunk\(\times\)dataset rather than assume one. Softmax stays optional for
UQ-oriented deployments.

\FloatBarrier
\section{Discussion}
\label{sec:discussion}

\paragraph{API surface.}
\textbf{Nimbus Personalizer} is a single contract: wrap any
frozen \texttt{encode}, personalize with a Bayesian head, optionally
escalate to affine-\(Z\), ship \texttt{BrainState}.
The primary contribution is that shared surface (systems), not a new
classifier family: OEM integrators integrate once and swap trunks instead
of standing up a fine-tune / PEFT pipeline per architecture.
In that light the head-family study of Section~\ref{sec:heads} is a
validation of API flexibility, not an ML-novelty claim: because head
optimality is dataset- and stress-dependent, the Personalizer must keep
heads modular (an interchangeable parameter) rather than hardcoded, with
LDA as the safe default. Head and cost cells show where the surface earns
its keep when capacity exists, and that cal-only is a cheap no-op when
clean already wins.

\paragraph{Downstream consumers of \texttt{BrainState}.}
The Personalizer stops at a structured prediction object: primary intent,
normalized uncertainty, and ranked alternatives---not a bare label.
The uncertainty is empirically calibrated where it matters most: under
severe shift the LDA head's expected calibration error falls well below
the frozen trunk's on both primary datasets (BNCI \(0.22\to 0.10\), Zhou
\(0.27\to 0.09\); EEGNet, \(n{=}40\)), so the \texttt{BrainState}
confidence field is a signal a downstream controller can act on rather
than an uninterpreted score. A future Decision Engine can treat those
fields as \emph{inputs} to observe--allocate--adapt loops without changing
the encoder--head contract.
We deliberately do not specify allocation rules here: the public core
exposes calibrated signals; control logic that spends resources remains a
separate layer~\citep{musienko2026ladder}.

\paragraph{Runtime and storage cost of the head.}
The cost story is not only adaptation fit time (Section~\ref{sec:cost}).
At inference, the Personalizer head adds a single matrix operation on top
of the frozen forward pass: measured per-trial latency on an Apple MPS
device is \(2.1\)\,ms p50 for EEGNet and \(7.6\)\,ms for the larger
Shallow trunk (\(n{=}9\), BNCI), well within a real-time BCI loop.
On the storage side, personalization under full fine-tune checkpoints the
\emph{entire} trunk per user, whereas the Personalizer persists only the
head's sufficient statistics: on foundation REVE the trainable-parameter
contrast is stark---full fine-tune updates \({\sim}6.9{\times}10^{7}\)
parameters, point LoRA \({\sim}6.1{\times}10^{4}\), last-layer
\({\sim}1.5{\times}10^{3}\), and the Bayesian head zero (it stores only
class means and covariance). Per-user state therefore scales with
embedding dimension, not trunk size, which is what makes the head the
deployable default for a multi-user OEM surface.

\paragraph{Foundation trunks.}
Larger frozen foundation encoders fit the same design insofar as they
expose a frozen trial-to-embedding map.
On pretrained REVE the Personalizer remains far cheaper than short
last-layer / full fine-tune; point LoRA is the PEFT intermediate method
that is unfair to reject from EEGNet-only walls (Figure~\ref{fig:reve-ft}).
Absolute REVE accuracy stays below cal-trained classical trunks; the
thesis is one API with a cheap personalization mid-point, not
foundation SOTA. We note for the BCI reader that classical Riemannian
pipelines (e.g., tangent-space LDA on signal covariances) give strong
session transfer on raw EEG but cannot be applied directly to the frozen
embedding spaces of deep or foundation trunks; the Personalizer head is
what bridges deep representation learning with instant Bayesian
calibration on those embeddings.

\paragraph{Other synthetic stresses (supporting, not remapped).}
\label{par:other-stress}
The multi-trunk severity map and cost figures in this paper use
\textbf{one} synthetic operator family---channel mixing---plus natural
session structure as a clean/gate check, which keeps the ladder
comparable across trunks.
Earlier BNCI2014-004 EEGNet recovery cells under a \emph{different}
protocol (LOSO factory, same-day cal/eval, \(n{=}20\)) also stress
one-channel dropout and mild channel mixing; because that timing harness
differs from the ladder suite, we report those cells separately in
Appendix~\ref{app:other-stress} rather than merge them with the
ladder-transfer ratios here. Physical electrode displacement and clinical
SNR remain untested.

\paragraph{Scope and limitations.}
Our evaluation is scoped to specific boundaries, stated positively.
\begin{itemize}[leftmargin=1.4em,itemsep=0.15em]
\item \emph{Ladder schedule.} The ordinal severity\(\to\)level pattern is
  conditional rather than law-like: strict ordinal escalation holds on
  \(5/18\) cells and severe\(\to\)L2 on \(9/18\), so we report
  calibration-only-when-clean (\(12/18\)) as the reliable default and the
  stricter schedule as dataset-dependent.
\item \emph{Stress coverage.} The main severity sweep uses synthetic
  channel mixing for cross-trunk comparability; physical electrode
  displacement and clinical SNR remain for future work, and the
  Table~\ref{tab:other-stress} supporting cells are reported as a separate
  protocol rather than a remapping of the ladder.
\item \emph{Control layer.} Allocation rules that spend labels or compute
  are out of scope here and treated in the companion control-layer paper.
\end{itemize}
The head-dominance boundaries (L1 does not universally beat L0; no single
best head; intermediate PEFT methods stay per-trunk spotlights) are
consolidated in the \S\ref{sec:heads} boundaries paragraph.

\paragraph{Statistics.}
All cells are subject means over small samples: BNCI \(n{=}9\), Zhou
\(n{=}4\), Kumar \(n{=}18\), Shin \(n{=}29\) (budget clamped to
\(n{=}20\)). Subject-level paired-bootstrap 95\% CIs
(\S\ref{sec:heads}; \(n_{\mathrm{boot}}{=}2000\), seed 0) average repeated
trunk measurements within participant before resampling participants. The
L1-over-L0 gain is nonzero only on Zhou and the QDA-over-LDA gain only on
BNCI; the other intervals span zero.
Kumar online sessions differ from offline MI; the REVE cost bake-off
covers Zhou and Kumar only (BNCI is 3-channel and stays near chance under the
same contract). All results are exploratory; no confirmatory tests were
pre-registered.

\FloatBarrier
\section{Conclusion}
\label{sec:conclusion}

Nimbus Personalizer's contribution is a trunk-agnostic contract---wrap,
Bayesian head, optional affine recovery, \texttt{BrainState}---that runs
across five heterogeneous classical trunks and four datasets, and extends
to Hub REVE under the same surface without a new personalization stack.
Where capacity exists, that surface is a cheap mid-point versus
warm-start FT / PEFT; where clean already wins, cal-only is the right
no-op (\(12/18\)).
Strict ordinal escalation is conditional (\(5/18\)), not the product
headline.
Ship the Personalizer API as the default personalization path; keep
fine-tune / PEFT as priced escape hatches; let companion work decide
\emph{when} to spend.

\section*{Acknowledgments}
Exploratory multi-trunk results were produced with an internal benchmark
suite accompanying Nimbus Personalizer; the suite is not required to use
the Personalizer API.

\FloatBarrier
\clearpage
\bibliographystyle{plainnat}
\bibliography{references}

\clearpage
\appendix

\section{Cross-stress replication of the head family}
\label{app:cross-stress}

The main paper scopes the head-family result (Section~\ref{sec:heads}) to
the synthetic channel-mix operator used throughout. To check whether the
QDA-dominance finding is an artifact of that one operator or a property
of the head, we re-ran the same LDA/QDA/Softmax head bake-off under three
further synthetic stress families---amplitude scaling, channel dropout,
and SNR noise---on the same five classical trunks $\times$ three datasets
(BNCI, Zhou, Kumar; 60 cells at severity $0.75$, $n{=}40$). This appendix
reports that sweep in full so a reader can judge the head family across
artifact types, not only channel mixing.

\paragraph{Protocol.}
Each stress family perturbs the raw multi-channel evaluation EEG (not the
embeddings) with the same calibration/adaptation/evaluation geometry and
the same frozen trunks as the main paper. Amplitude scaling multiplies
each channel by a subject-specific factor; channel dropout zeros a random
channel; SNR noise adds white Gaussian noise to reach a target SNR. All
are applied at the severe level; calibration stays clean. Utility uses the
shared definition~\eqref{eq:U}; subject-level means and percentile
bootstrap 95\% CIs use $n_{\mathrm{boot}}{=}2000$, seed~0.

\paragraph{Does QDA beat LDA under each stress?}
Table~\ref{tab:cross-stress-pooled} pools the per-subject QDA-minus-LDA
utility delta across the field. The effect is near zero or negative under
three of the four families; only channel mixing (the main-paper operator)
and feature shift trend positive, and only amplitude scaling's CI
excludes zero (in the wrong direction). QDA dominance is therefore not a
stress-family-independent property of the head.

\begin{table}[H]
\centering
\caption{Cross-stress pooled QDA-minus-LDA utility delta (60-cell field,
5 trunks $\times$ 3 datasets, severity $0.75$). Percentile bootstrap
95\% CI; $n_{\mathrm{boot}}{=}2000$, seed~0. ``CI $\ni 0$'' = interval
includes zero. Channel mixing is the main-paper operator, reported here
for reference.}
\label{tab:cross-stress-pooled}
\small
\begin{tabular}{lcccc}
\toprule
Stress family & Pooled $\Delta U$ (QDA$-$LDA) & 95\% CI & $n$ cells & CI $\ni 0$ \\
\midrule
Amplitude scaling & \(-0.015\) & \([-0.027,-0.004]\) & 15 & no (QDA worse) \\
Channel dropout   & \(-0.006\) & \([-0.017,+0.005]\) & 15 & yes \\
Feature shift     & \(+0.007\) & \([-0.008,+0.021]\) & 15 & yes \\
SNR noise         & \(+0.001\) & \([-0.009,+0.011]\) & 15 & yes \\
\bottomrule
\end{tabular}
\end{table}

\begin{figure}[H]
\centering
\includegraphics[width=0.96\linewidth]{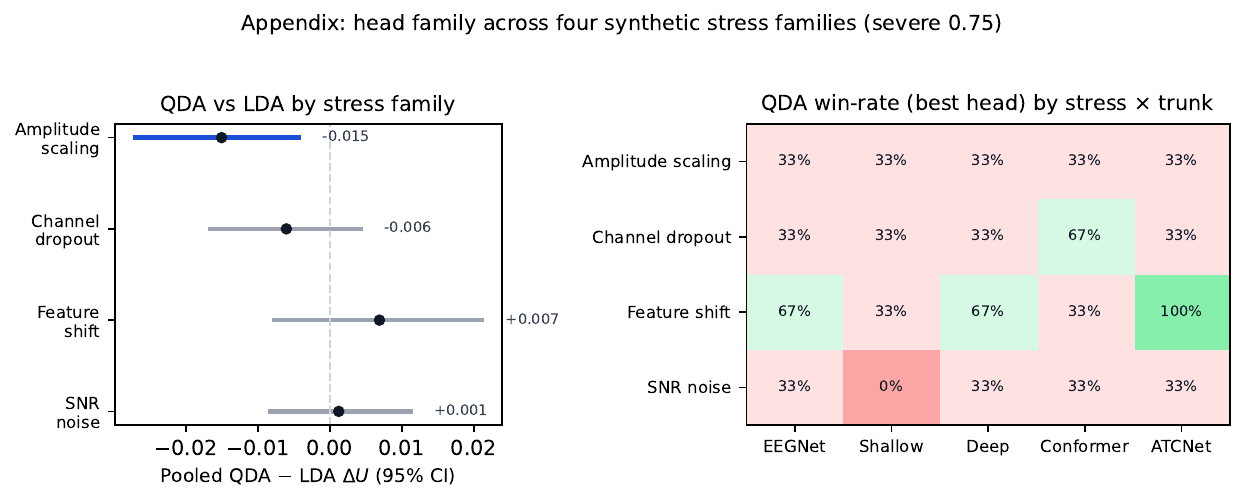}
\caption{Cross-stress head-family summary (severity \(0.75\)).
Left: pooled QDA-minus-LDA \(\Delta U\) with 95\% bootstrap CI per stress
family (blue where the CI excludes zero).
Right: fraction of cells where QDA is the best head, by
stress\(\times\)trunk (60 cells). QDA wins are sparse and
stress-dependent; no family produces QDA dominance across trunks.}
\label{fig:cross-stress}
\end{figure}

\paragraph{Where does QDA still win?}
Figure~\ref{fig:cross-stress} maps the win-rate (cells where QDA is the
best Personalizer head) by stress family and trunk. Even in the most
QDA-favorable family (feature shift), QDA wins only \(9/15\) cells, and
the per-trunk win-rate never exceeds chance by a wide margin. The
takeaway is consistent with the main paper: LDA is the safe default and
QDA is a per-cell option to benchmark, not a universally preferred head.

\paragraph{Interpretation.}
The cross-stress sweep closes the ``what about other artifacts?'' question
empirically: the head family is robust as an \emph{API} (it runs and
recovers under all four families) but the \emph{ranking} of heads is not.
Physical electrode displacement and clinical SNR under real recording
conditions remain for future work; the four synthetic families here are a
controlled proxy, and they already show that any universal head-choice
claim would be premature.

\section{Supporting recovery cells under a separate protocol}
\label{app:other-stress}

These BNCI2014-004 EEGNet cells use a \emph{different} protocol than the
18-cell ladder map: a leave-one-subject-out factory decoder, same-day
calibration/evaluation, and a labeled budget of \(n{=}20\) rather than
\(n{=}40\). They corroborate that the Personalizer recovers cheaply when
the factory decoder is gated below \(0.70\), but because the timing
harness differs from the ladder suite their fit-wall ratios
(\({\sim}30\times\)) are \textbf{not} comparable to, and must not be
merged with, the ladder-transfer CUDA L3/L1 ratios reported in the main
text. We list them here for completeness, not as a remapping of the main
severity sweep.

\begin{table}[H]
\centering
\caption{Supporting BNCI2014-004 EEGNet cells under a separate protocol
(LOSO factory, same-day cal/eval, \(n{=}20\)).
\(\Delta\mathrm{acc}\) is Personalizer minus frozen; wall ratio is
fine-tune / Personalizer adapt fit. Exploratory.}
\label{tab:other-stress}
\small
\begin{tabular}{lcccc}
\toprule
Synthetic stress & Frozen acc & Pers.\ acc & \(\Delta\mathrm{acc}\) & Wall FT/Pers. \\
\midrule
1-channel dropout & 0.654 & 0.740 & \(+0.086\) & \({\sim}30\times\) \\
Mild channel mix & 0.542 & 0.756 & \(+0.214\) & \({\sim}34\times\) \\
\bottomrule
\end{tabular}
\end{table}

\end{document}